\documentclass[superscriptaddress,showpacs,amssymb,10pt,reprint,aps,prd,longbibliography,nofootinbib,floatfix]{revtex4-1}

\usepackage{graphicx,epsfig,amssymb}
\usepackage{amsmath,amsfonts, times}
\usepackage{bm} 

\usepackage{epstopdf}
\usepackage[linktocpage,colorlinks]{hyperref}
\usepackage[caption=false]{subfig}
\usepackage[usenames]{color}     
\usepackage{natbib}
\usepackage{soul}
\usepackage[utf8x]{inputenc}
\usepackage[usenames]{color}

\definecolor{coolblack}{rgb}{0.0, 0.18, 0.39}
\definecolor{darkred}{rgb}{0.5,0,0}
\definecolor{darkgreen}{rgb}{0,0.5,0}
\definecolor{darkblue}{rgb}{0,0,0.5}
\definecolor{lapislazuli}{rgb}{0.15, 0.38, 0.61}
\definecolor{venetianred}{rgb}{0.78, 0.03, 0.08}
\definecolor{bleudefrance}{rgb}{0.19, 0.55, 0.91}
\definecolor{dogwoodrose}{rgb}{0.84, 0.09, 0.41}
\hypersetup{colorlinks=true, citecolor=darkgreen, linkcolor=darkblue, 
	urlcolor = blue}

\newcommand\numberthis{\addtocounter{equation}{1}\tag{\theequation}}

\def\be{\begin{equation}}
\def\ee{\end{equation}}

\newcommand{\bea}{\begin{eqnarray}}
\newcommand{\eea}{\end{eqnarray}}
\newcommand{\ben}{\begin{enumerate}}
	\newcommand{\een}{\end{enumerate}}
\newcommand{\bi}{\begin{itemize}}
	\newcommand{\ei}{\end{itemize}}

\newcommand{\rt}{r_{\star}}

\def\ga{\mathrel{\raise.3ex\hbox{$>$\kern-.75em\lower1ex\hbox{$\sim$}}}}
\def\la{\mathrel{\raise.3ex\hbox{$<$\kern-.75em\lower1ex\hbox{$\sim$}}}}

\def\l{\left}
\def\r{\right}
\def\be{\begin{equation}}
\def\ee{\end{equation}}

\def\I_M{{I_{\scriptscriptstyle M\times M}}}

\def\be{\begin{equation}}
\def\ee{\end{equation}}
\def\bea{\begin{eqnarray}}
\def\eea{\end{eqnarray}}
\newcommand{\beq}{\begin{eqnarray}}
\newcommand{\eeq}{\end{eqnarray}}
\def\pa{\partial}

\begin{document}
	\title{\large {Absorption by deformed black holes}}

	\author{Renan B. Magalh\~aes}
	\email{rbmagalhaes22@hotmail.com}
	\affiliation{Programa de P\'os-Gradua\c{c}\~{a}o em F\'{\i}sica, Universidade 
		Federal do Par\'a, 66075-110, Bel\'em, Par\'a, Brazil.}
	
	\author{Luiz C. S. Leite}
	\email{luizcsleite@ufpa.br}
	\affiliation{Programa de P\'os-Gradua\c{c}\~{a}o em F\'{\i}sica, Universidade 
		Federal do Par\'a, 66075-110, Bel\'em, Par\'a, Brazil.}
	\affiliation{Campus Altamira, Instituto Federal do Par\'a, 68377-630, Altamira, Par\'a, Brazil.}
	
	\author{Lu\'is C. B. Crispino}
	\email{crispino@ufpa.br}
	\affiliation{Programa de P\'os-Gradua\c{c}\~{a}o em F\'{\i}sica, Universidade 
		Federal do Par\'a, 66075-110, Bel\'em, Par\'a, Brazil.}

	\begin{abstract}
	Alternative theories of gravity and the parameterized deviation approach allow black hole solutions to have additional parameters beyond mass, charge and angular momentum. Matter fields could be, in principle, affected by the additional parameters of these solutions. 
We compute the absorption cross section of massless spin-0 waves by static Konoplya-Zhidenko black holes, characterized by a deformation parameter introduced in the mass term, and compare it with the well-known absorption of a Schwarzschild black hole with the same mass. We compare our numerical results with the sinc approximation in the high-frequency limit, finding excellent agreement. 
	\end{abstract}
	
	\date{\today}
	
	\maketitle

\section{Introduction}	

Among General Relativity (GR) predictions, stands out one of the most fascinating objects of Modern Physics --- the black holes (BHs) --- which are described by a one-way membrane, called event horizon. BH physics has been based in a set of theorems that establishes their simplicity --- the no-hair theorems~\cite{israel,carter,robinson}. Along the years, experimental efforts have been made to test the validity of the no-hair theorems~\cite{gossan,meidan,isi}, and, so far, no experimental data has contradicted the no-hair paradigm.

Regardless the excellent agreement of GR with experiments~\cite{eht,ligo}, some extensions of GR have been proposed as alternative theories of gravity, with their corresponding BH solutions. In this context, it is interesting to compare each of these modified BH solutions with the well-known BHs of GR. Moreover, BHs could be described by an unknown gravity theory, and recently Johannsen and Psaltis introduced a parametric deviation approach~\cite{johannsen}, without specifying the dynamical equations which would replace the Einstein ones. This strategy avoids, for instance, some limitations of the original bumpy BH approach~\cite{collins,vigeland,yunes}. Among the parametric deformed solutions that emerged, Konoplya and Zhidenko proposed a Kerr-like solution, introducing a parametric deformation in the mass term, keeping the asymptotic behavior of the Kerr spacetime, but changing how the mass of the BH influences the event horizon vicinity~\cite{KZ}.

Among the developments in the study of fields in the vicinity of BHs, the absorption of waves plays an important role~\cite{fabbri,unruh,sanchez}. It is well-known that the absorption process is related to the BH parameters, namely the mass, electric charge and angular momentum~\cite{BODC:2014,MC:2014,CDHO:2015,BC:2016,LDC:2017,LDC:2018,LBC:2019}. However, deformed BH solutions present additional parameters, that can, in principle, also modify the absorption process. The investigation of how the additional parameters influence the absorption, enables us to better understand the interactions of the deformed BH with fields, allowing us to compare them with the GR BH solutions.

In this letter we take a first step in the analysis of the absorption of waves by deformed BHs, considering the static version of the Konoplya-Zhidenko BH (KZBH), aiming to understand the role played by the deformation parameter in the absorption process.
We compute the scalar absorption by the static Konoplya-Zhidenko BH (SKZBH).
The remaining of this letter is organized as follows. In Sec.~\ref{sec:KZ} we present the SKZBH and explore some properties of this spacetime. In Sec.~\ref{sec:planar} we consider the massless spin-0 field in the surroundings of a SKZBH and study its dynamics. In Sec.~\ref{sec:absorption} we analyze the absorption cross section of the massless scalar field. In Subsec.~\ref{sec:absorption1} we briefly discuss the numerical approach we used to obtain the absorption cross section, and show our results in Subsec.~\ref{sec:absorption2}. We conclude with our final remarks in Sec.~\ref{sec:final}. We use the natural units $G=c=\hbar=1$, and signature $(+--\,-)$.

\section{Static Konoplya-Zhidenko black hole}\label{sec:KZ}

Let us first consider a Schwarzschild BH line element with mass $M$, namely 
\be
\label{eq:schw}
ds^2_{s}=f_{s}(r)dt^2-\frac{1}{f_{s}(r)}dr^2-r^2d\Omega^2,
\ee
where $f_{s}(r) \equiv 1- 2M/r$ and  $d\Omega^2\equiv d\theta^2+\sin^2\theta d\phi^2$ is the line element of a unit sphere. We can introduce some parametric deformations in the mass term $M$~\cite{KZ}, as a power series of $1/r$, namely
\be
\label{eq:deformation_T1}
M \rightarrow M + \dfrac{1}{2}\sum_{i=0}^{\infty}\dfrac{\eta_{i}}{r^{i}}.
\ee

This changes the $f_{s}(r)$ function as
\be
\label{eq:KZmetricfunctions_T1}
f_{s}(r)\rightarrow 1-\dfrac{2M}{r}-\sum_{i=0}^{\infty}\dfrac{\eta_{i}}{r^{i+1}},
\ee

so that the line element becomes
\begin{align*}
\label{eq:KZmetric_T1}
ds_{PD}^2=&\left(1-\dfrac{2M}{r}-\sum_{i=0}^{\infty}\dfrac{\eta_{i}}{r^{i+1}}\right)dt^2\\&-\left(1-\dfrac{2M}{r}-\sum_{i=0}^{\infty}\dfrac{\eta_{i}}{r^{i+1}}\right)^{-1}dr^2-r^2d\Omega^2. \numberthis
\end{align*}

Some constraints can be imposed to the parameters $\eta_{i}$. In order to do it, we look for the asymptotic limit of Eq.~\eqref{eq:KZmetric_T1}. Far from the central object, the line element reduces to
\begin{align*}
\label{eq:KZmetric_asymptotic_T1}
ds_{PD}^2\Big\vert_{r\, \gg \, 2M}&\approx \left(1-\dfrac{2M}{r}-\dfrac{\eta_{0}}{r}-\dfrac{\eta_{1}}{r^{2}}\right)dt^2+\\&-\left(1-\dfrac{2M}{r}-\dfrac{\eta_{0}}{r}-\dfrac{\eta_{1}}{r^{2}}\right)^{-1}dr^2-r^2d\Omega^2. \numberthis
\end{align*}

Therefore, if one seeks to obtain the same behavior of the Schwarzschild line element~\eqref{eq:schw} in the far field limit, the condition $\eta_{0}=0$ must hold. The constraint in the parameter $\eta_{1}$ is obtained by the parameterized post-Newtonian (PPN) approach~\cite{PPN}, for which the asymptotic spherically symmetric spacetime has the form~\cite{johannsen}
\begin{equation}
ds_{PPN}^{2}\Big\vert_{r\,\gg\,2M} = A(r)dt^2 - B(r)dr^2 -r^2d\Omega^2,
\end{equation}

where
\begin{align}
A(r) &\equiv 1 - \dfrac{2M}{r} + 2(\beta - \gamma)\dfrac{M^2}{r^2},\\
B(r) &\equiv 1+2\gamma\dfrac{M}{r},
\end{align}

with $\beta$ and $\gamma$ being PPN parameters.

From Cassini experiments~\cite{bertotti} and the experimental data of Lunar Laser Ranging~\cite{boggs} we have
\begin{equation}
\label{eq:constraint}
|\beta - \gamma|\leqslant 2.3 \times 10^{-4},
\end{equation} so that,
\begin{equation}
\label{eq:constraint_2}
\dfrac{\eta_{1}}{2}\leqslant 2.3 \times 10^{-4}.
\end{equation}
Hence, we assume that the term containing the deformation parameter $\eta_{1}$ is negligible and set $\eta_{1}=0$, so that the asymptotic form of the deformed line element reduces to the Schwarzschild line element. Taking into account the previous arguments, we keep a single deformation in the mass term $M$, following Konoplya and Zhidenko~\cite{KZ} procedure for the Kerr BH, changing $M$ as:
\be
\label{eq:deformation}
M \rightarrow M + \dfrac{\eta}{2\,r^2}.
\ee
We note that the parametrization given by Eq.~\eqref{eq:deformation} is equivalent to assume that $\eta_{i}=\eta\, \delta_{i\,2}$ in Eq.~\eqref{eq:deformation_T1}.

We call $\eta$ the Konoplya-Zhidenko (KZ) deformation parameter. With the change~\eqref{eq:deformation}, the $f(r)$ function becomes
\be
\label{eq:KZmetricfunctions}
f_{KZ}(r)\equiv 1-\dfrac{2M}{r}-\dfrac{\eta}{r^3},
\ee
associated to the line element of the SKZBHs, namely
\begin{align*}
\label{eq:KZmetric}
ds^2=&\left(1-\dfrac{2M}{r}-\dfrac{\eta}{r^3}\right)dt^2\\&-\left(1-\dfrac{2M}{r}-\dfrac{\eta}{r^3}\right)^{-1}dr^2-r^2d\Omega^2. \numberthis
\end{align*}
If the KZ deformation parameter vanishes~($\eta=0$), Eq.~\eqref{eq:KZmetric} reduces to the well-known Schwarzschild line element~\eqref{eq:schw}. It is important to notice that the line element~\eqref{eq:KZmetric} does not represent a solution of the Einstein's equations.

The event horizon location of the SKZBH, with deformation parameter $\eta$, can be obtained evaluating
\be
\label{eq:eventhorizon}
g^{rr} = f_{KZ}(r) = 0,
\ee
resulting in a cubic equation, having three different solutions $r_{i}$~\cite{long}, namely
\small
\begin{align}
\label{eq:r1}  r_{1}(\eta) &= \dfrac{1}{3}\left( 2M + \dfrac{4 \sqrt[3]{2}M^2}{A(\eta)} + \dfrac{A(\eta)}{\sqrt[3]{2}} \right),\\
\label{eq:r2} r_{2}(\eta) &= \dfrac{1}{3}\left( 2M - \dfrac{2 \sqrt[3]{2}(1 + i\sqrt{3})M^2}{A(\eta)} - \dfrac{(1 - i\sqrt{3})A(\eta)}{2\sqrt[3]{2}} \right),\\
\label{eq:r3} r_{3}(\eta) &= \dfrac{1}{3}\left( 2M - \dfrac{2 \sqrt[3]{2}(1 - i\sqrt{3})M^2}{A(\eta)} - \dfrac{(1 + i\sqrt{3})A(\eta)}{2\sqrt[3]{2}} \right),
\end{align}
\normalsize
where 
\begin{equation}
A(\eta) \equiv \sqrt[3]{16M^3 + 27 \eta + 3\sqrt{3}\sqrt{32M^3\eta+27\eta^2}}.
\label{eq:A_eta}
\end{equation}

\begin{figure}
\centering
\includegraphics[width=3.0in]{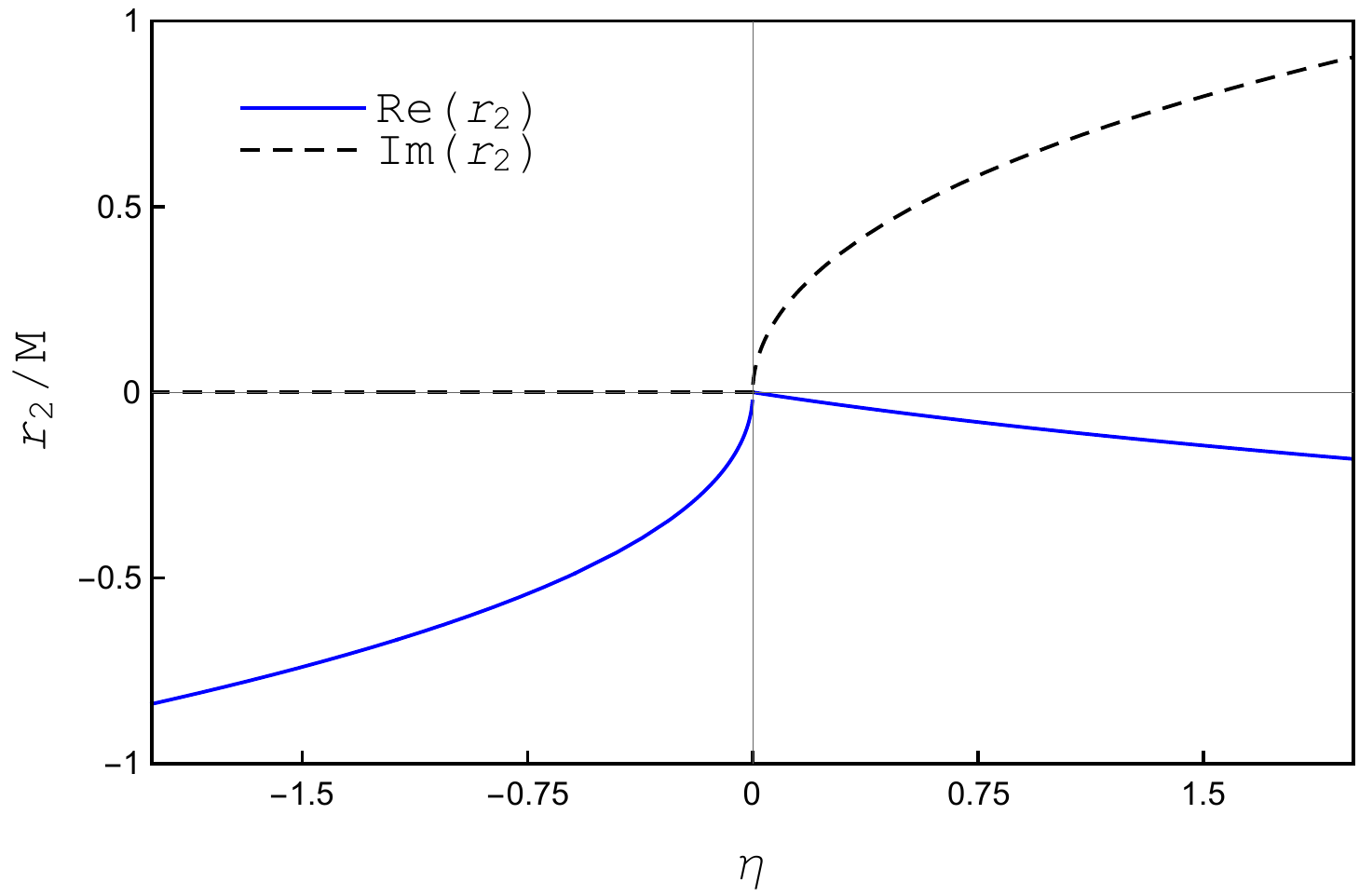}
\caption{The real and imaginary parts of the solution $r_{2}$, given by Eq.~\eqref{eq:r2}. For positive values of $\eta$, the solution has a non-vanishing imaginary part. Although for negative values of $\eta$ the solution $r_{2}$ has a vanishing imaginary part, the real part is negative, and therefore $r_{2}$ does not represent a radius.}
\label{fig:r2}
\end{figure}

From Fig.~\ref{fig:r2} we see that when the imaginary part of $r_{2}(\eta)$ vanishes, its real part is negative. Therefore, the solution $r_{2}(\eta)$ is not associated to an event horizon.

\begin{figure}
\centering
\includegraphics[width=3.0in]{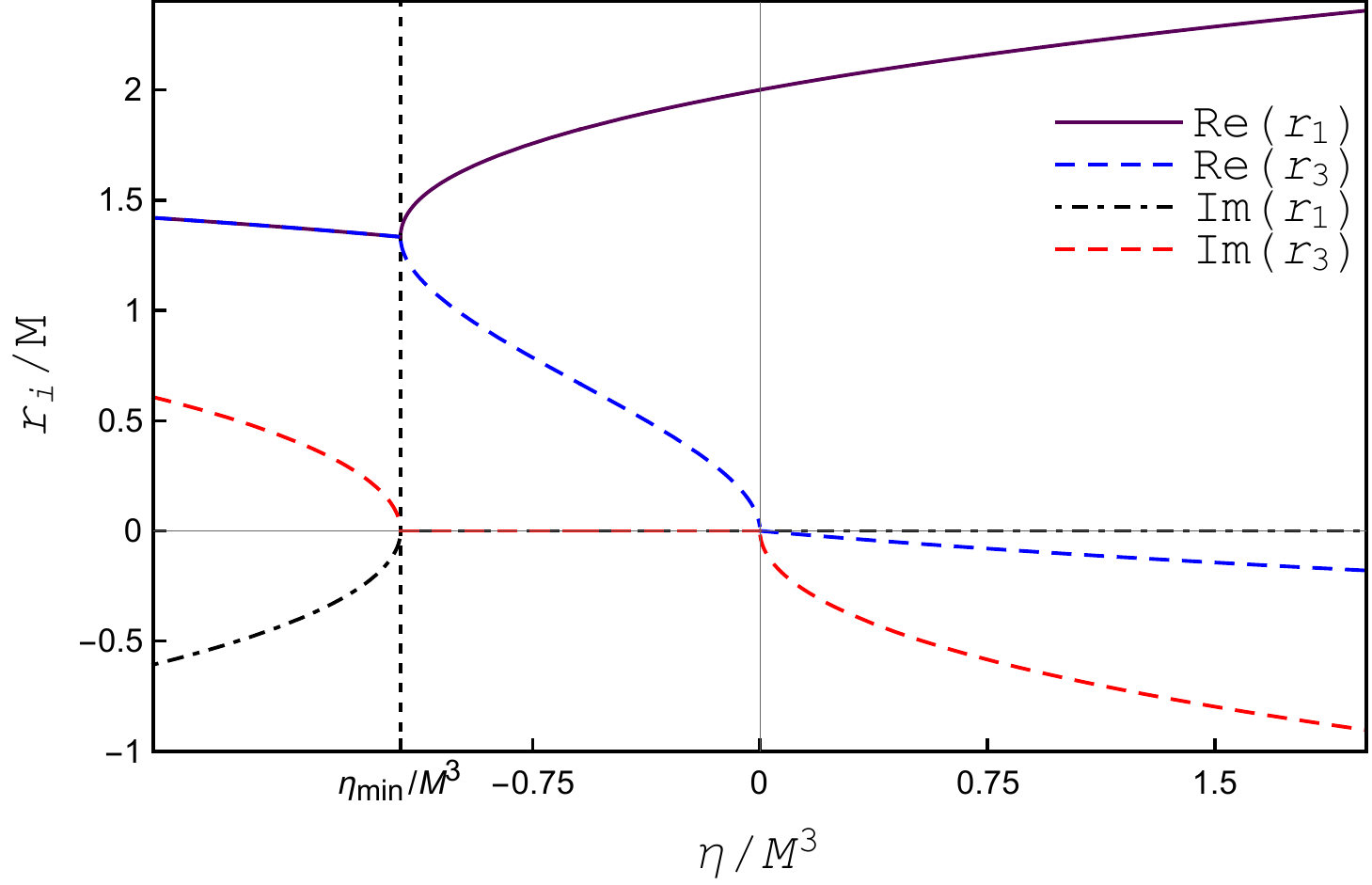}
\caption{The real and imaginary parts of the solutions $r_{1}$ and $r_{3}$, given by Eqs.~\eqref{eq:r1} and~\eqref{eq:r3}, respectively. For positive values of $\eta$, the solution $r_{1}$ has a vanishing imaginary part, while the imaginary part of the solution $r_{3}$ is non-vanishing. For $\eta_{\text{min}} < \eta < 0$, $r_{1}$ and $r_{3}$ have vanishing imaginary parts, so that the SKZBHs have two horizons, the event horizon being the external one. For $\eta<\eta_{\text{min}}$, the imaginary parts of both $r_{1}$ and $r_{3}$ are non-vanishing. The dashed vertical line represents $\eta=\eta_{\text{min}}$.}
\label{fig:r1r3}
\end{figure}

From Fig.~\ref{fig:r1r3} we see that when the imaginary parts of  $r_{1}(\eta)$ and $r_{3}(\eta)$ vanish, their respective real parts are non-negative. There is a value of the deformation parameter, $\eta_{\text{min}}=-32/27M^3$, such that, for $\eta<\eta_{\text{min}}$ the imaginary parts of both $r_{1}(\eta)$ and $r_{3}(\eta)$ become non-zero.
Hence, for $\eta<\eta_{\text{min}}$, the static KZ spacetime represents a naked singularity, violating the cosmic censorship conjecture~\cite{penrose}, and we shall not consider it here. For values of the deformation parameter in the range $-32/27M^3\leq \eta<0$, the radii $r_{1}(\eta)$ and $r_{3}(\eta)$ represent two horizons of the SKZBH, with the outer radius, $r_{1}(\eta)$, being the event horizon. For positive values of $\eta$, the imaginary part of $r_{3}(\eta)$ is non-zero, so that the only horizon of the SKZBH, in this range, is the radius $r_{1}$. From here on we will label $r_{1}(\eta)\equiv r_{h}(\eta)$ and assume $\eta\geq\eta_{\text{min}}$.

We conclude that the KZ deformation parameter $\eta$ influences in the event horizon position, allowing SKZBHs with the same ADM mass $M$ to have different horizon radius $r_{h}$.

\begin{figure}
\centering
\includegraphics[width=3.0in]{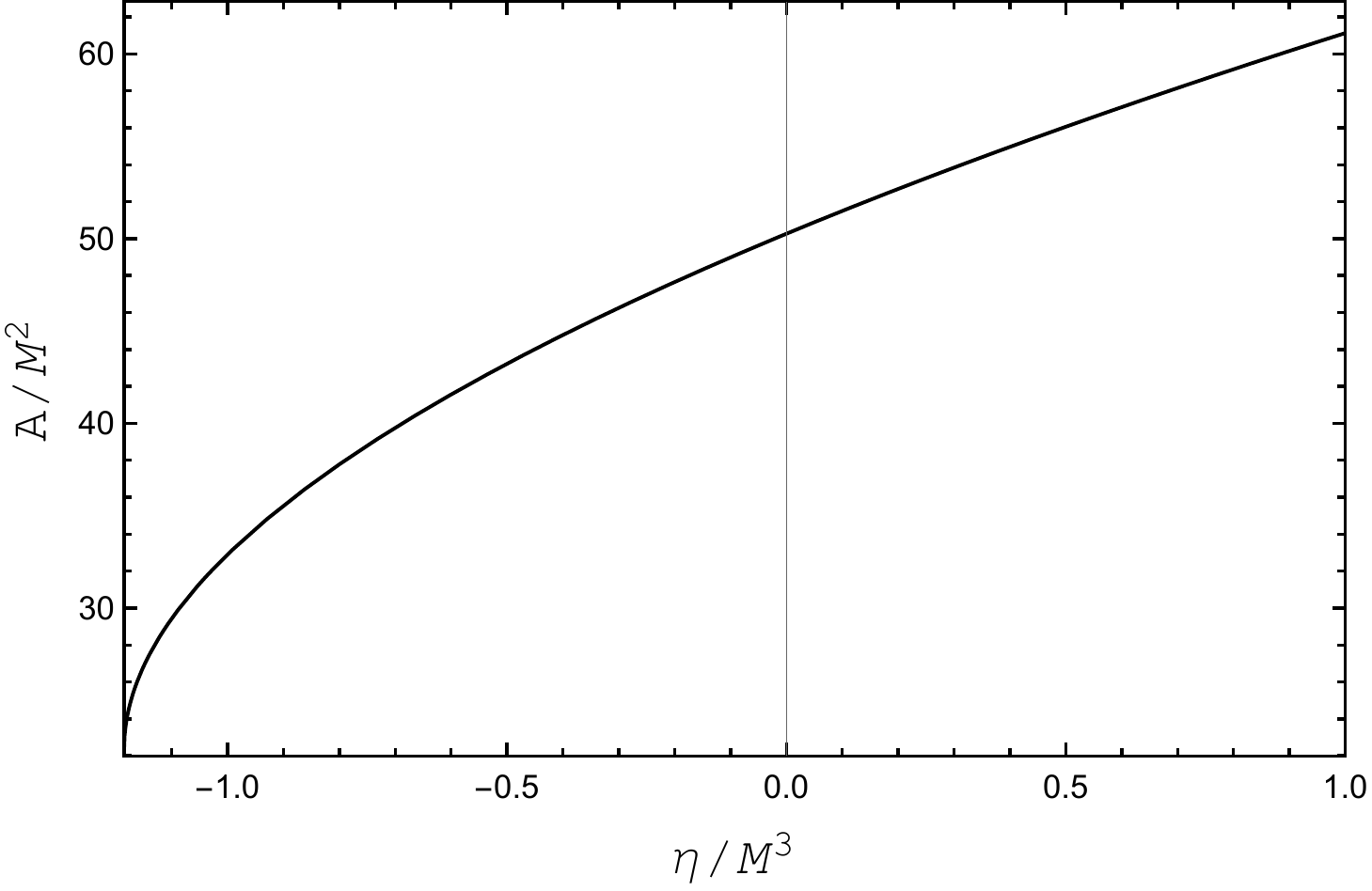}
\caption{The surface area of the SKZBH, given by Eq.~\eqref{eq:Area_eta}, as a function of the deformation parameter $\eta$.}
\label{fig:area}
\end{figure}

In Fig.~\ref{fig:area} we plot the event horizon area of the SKZBH,
\begin{equation}
A(\eta)=4\pi r^2_{h}(\eta),
\label{eq:Area_eta}
\end{equation}
and, we see that if we increase the deformation parameter $\eta$ the surface area of the SKZBH increases.

\section{Scalar waves in the SKZBH spacetime}{\label{sec:planar}}
We consider a massless spin-0 field $\Psi$, with dynamics governed by the Klein-Gordon equation, namely
\be
(-g)^{-1/2}\pa_\mu(g^{\mu\nu}\sqrt{-g}\pa_\nu\Psi)=0,\label{eq:kge}
\ee
where $\pa_\mu\equiv\pa/\pa x^\mu$, and $g$ is the determinant of the metric with contravariant components $g^{\mu\nu}$. One can separate the partial differential equation~\eqref{eq:kge} in a set of ordinary differential equations, by decomposing the massless spin-0 field~$\Psi$ as follows:
\be
\Psi(x^\mu)=\dfrac{\psi_{\omega l}(r)}{r}S_{lm}(\theta,\,\phi)e^{-i\omega t},
\label{eq:decom}
\ee
where $\omega$ is the frequency of the scalar field. Substituting Eq.~\eqref{eq:decom} into Eq.~\eqref{eq:kge}, we obtain
\begin{align*}
&\dfrac{r^2}{f(r)}\left\{\omega^2 \psi_{\omega l} + f(r)\frac{d}{dr}\l[f(r)\frac{d\psi_{\omega l}}{dr}\r] - \dfrac{f(r)}{r}\dfrac{df}{dr} \psi_{\omega l}\right\} =\\ &-\dfrac{1}{S_{lm}}\l[\dfrac{\partial^2S_{lm}}{\partial \theta^2} + \cot \theta\dfrac{\partial S_{lm}}{\partial \theta} + \dfrac{1}{\sin^2\theta}\dfrac{\partial^2 S_{lm}}{\partial \phi^2} \r]\psi_{\omega l}.
\numberthis \label{eq:kgs}
\end{align*}
Hence, as a consequence of the spherical symmetry of the spacetime, the angular dependence of $\Psi(x^\mu)$ is given by the scalar spherical harmonics, i.e. $S_{lm}(\theta,\phi)= Y_{lm}(\theta,\phi)$, which obey the following eigenvalue equation~\cite{zwillinger}
\be
\dfrac{\partial^2Y_{lm}}{\partial \theta^2} + \cot \theta\dfrac{\partial Y_{lm}}{\partial \theta} + \dfrac{1}{\sin^2\theta}\dfrac{\partial^2 Y_{lm}}{\partial \phi^2} = -l(l+1)Y_{lm}.
\label{eq:she}
\ee 
From Eqs.~\eqref{eq:kgs} and~\eqref{eq:she}, we obtain the ordinary differential equation for the radial function $\psi_{\omega l}(r)$, namely
\be
f(r)\frac{d}{dr}\l[f(r)\frac{d\psi_{\omega l}}{dr}\r]+\l[\omega^2-V_l(r)\r]\psi_{\omega l}=0,\label{eq:radialeq}
\ee
where $V_l$ is the effective potential, given by
\be
V_l(r)\equiv \frac{f(r)}{r}\frac{df}{dr}+f(r)\frac{l(l+1)}{r^2}. \label{eq:potential}
\ee

By introducing the tortoise coordinate $\rt$,
\be
\rt\equiv\int \frac{dr}{f(r)},
\ee
defined in the range $-\infty<\rt<+\infty$, we can rewrite Eq.~\eqref{eq:radialeq} as a Schrödinger-like equation, namely
\be
\frac{d^2\psi_{\omega l}}{d\rt^2}+\l[\omega^2-V_l\r]\psi_{\omega l}(\rt)=0.\label{eq:radialeqtor}
\ee
By analyzing the potential $V_{l}(r)$, we can obtain the asymptotic behavior of the radial solution when the field is close enough to the event horizon and when the field is far away from the BH.

\begin{figure}
\centering
\includegraphics[width=3.0in]{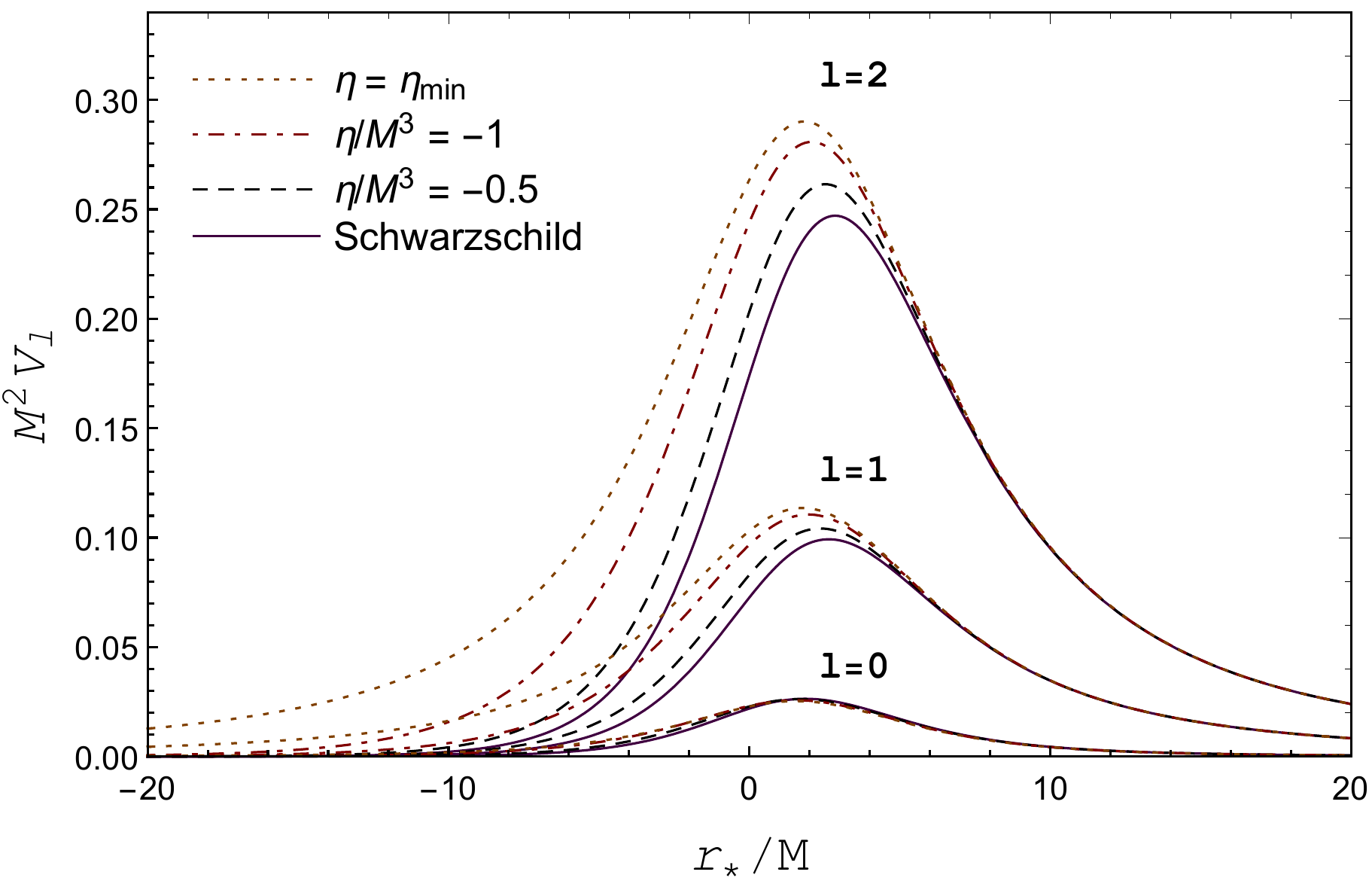}
\includegraphics[width=3.0in]{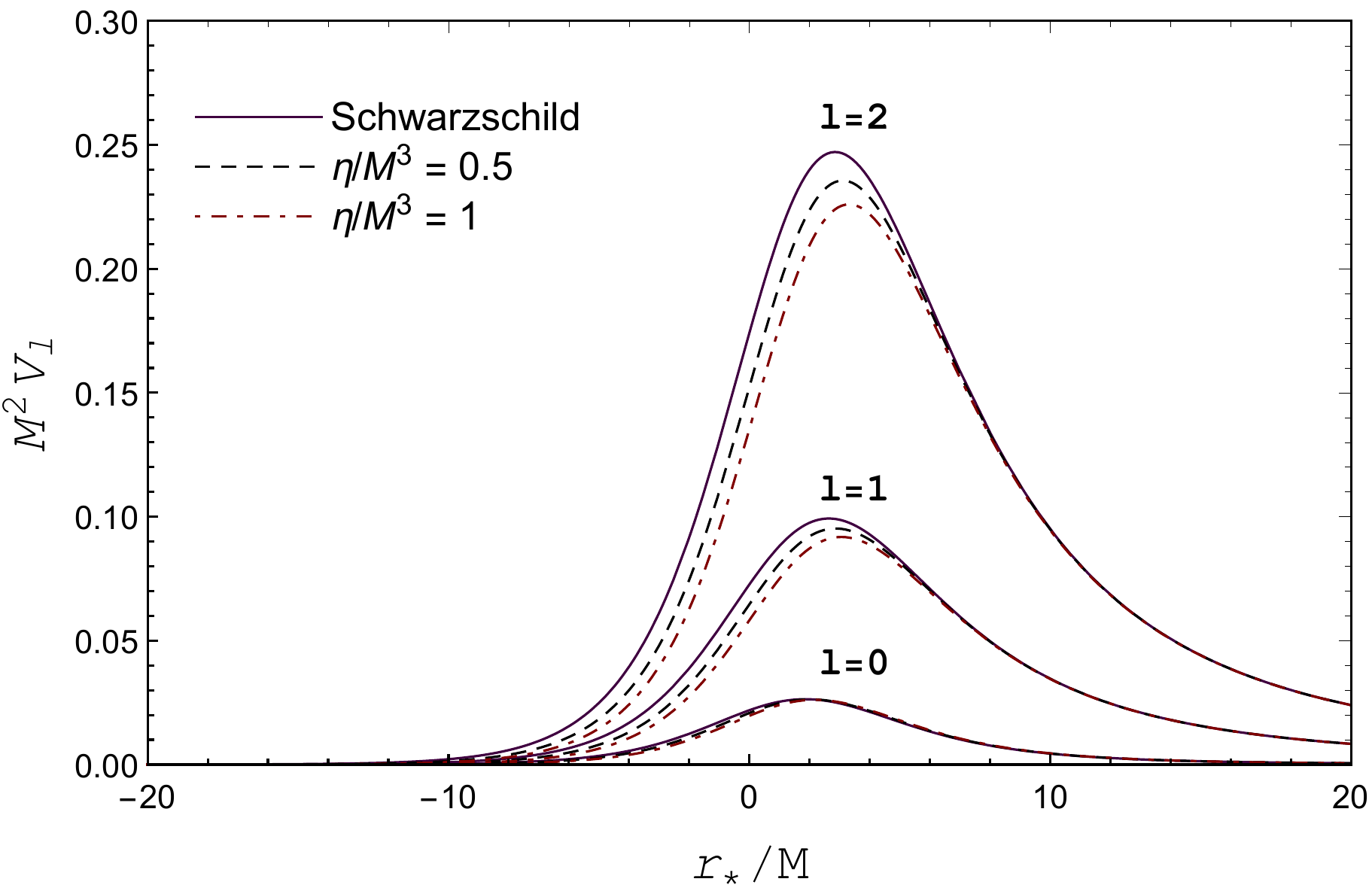}
\caption{The effective potential of the SKZBH with $l=0$, $1$ and $2$, for non-positive (top panel) and non-negative (bottom panel) values of the parameter $\eta$.}
\label{fig:potnonneg}
\end{figure}

In Fig.~\ref{fig:potnonneg} we plot the effective potential~\eqref{eq:potential} for non-positive (top panel) and non-negative (bottom panel) values of the deformation parameter $\eta$. We notice that as we increase the value of $\eta$, the peak of the effective potential decreases. We also notice that close to the event horizon $(\rt\rightarrow-\infty)$ and at the spacial infinity $(\rt\rightarrow\infty)$, the effective potential vanishes, regardless the value of the KZ deformation parameter $\eta$.

In these asymptotic limits ($r\rightarrow\infty$ and $r\rightarrow r_{h}$), the radial equation~\eqref{eq:radialeqtor} takes the form of a simple harmonic oscillator equation, namely
\begin{equation}
\label{eq:radialOHS1}\frac{d^2\psi_{\omega l}}{d\rt^2}\Big\rvert_{\rt \rightarrow -\infty}+\omega^2\psi_{\omega l}(\rt)\Big\rvert_{\rt \rightarrow -\infty}=0,
\end{equation}
and
\begin{equation}
\label{eq:radialOHS2}\frac{d^2\psi_{\omega l}}{d\rt^2}\Big\rvert_{\rt \rightarrow \infty}+\omega^2\psi_{\omega l}(\rt)\Big\rvert_{\rt \rightarrow \infty}=0.
\end{equation}
The solutions of Eqs.~\eqref{eq:radialOHS1} and~\eqref{eq:radialOHS2} are ingoing and outgoing waves.
For our scattering problem, we consider a wave incoming from infinity, which is partially transmitted through the potential barrier, being absorbed by the BH, and partially reflected back to infinity. This scenario corresponds to the following boundary conditions:   
\be
\psi_{\omega l}(\rt)\sim \l\{
\begin{array}{ll}
	A_{in}e^{-i \omega \rt}+A_{out}e^{i \omega \rt},&\rt\to+\infty, \\
	e^{-i\omega \rt},& \rt\to-\infty.
\end{array}\r.\label{eq:inmodes}
\ee

The reflection and transmission coefficients are related to ${\cal R}_{\omega l } = A_{out}/A_{in}$ and ${\cal T}_{\omega l } = 1/A_{in}$, respectively. Moreover, using flux conservation, it is possible to show that they satisfy the following relation:
\be
|{\cal R}_{\omega l }|^2+|{\cal T}_{\omega l }|^2=1.
\label{eq:RTrelation}
\ee
\section{Absorption}\label{sec:absorption}

A common strategy to compute the absorption cross section of asymptotically flat static BHs consists in considering the partial-wave approach, expanding an asymptotic monochromatic plane wave propagating along the $z$-axis, as
\be
e^{-i\omega(t - z)} = e^{-i\omega t}\sum^{\infty}_{l=0}(2l+1)i^{l}j_{l}(\omega r)P_{l}(\cos\theta),
\label{eq:partialwaves}
\ee
where $j_{l}(\omega r)$ are the spherical Bessel functions and $P_{l}(\cos\theta)$ are the Legendre polynomials. Asymptotically, the plane wave can be written as~\cite{BODC:2014}
\begin{align*}
e^{-i\omega(t - z)}\Big\vert_{r\rightarrow\infty} = \dfrac{e^{-i\omega t}}{2i\omega r}\sum^{\infty}_{l=0}(2l+1)(-1)^{l+1}\,\, \times \\ \l[e^{-i\omega r} + (-1)^{l+1}e^{i\omega r}\r]P_{l}(\cos\theta). \numberthis
\label{eq:partialwavesinf}
\end{align*}
Taking into account the expansion~\eqref{eq:partialwavesinf}, we may write a scalar wave propagating in an asymptotically flat spacetime as
\be
\phi_{\omega l} = \dfrac{e^{-i\omega t}}{2i\omega r}\sum^{\infty}_{l=0}(2l+1)(-1)^{l+1}\psi_{\omega l}(r)P_{l}(\cos\theta).
\label{eq:partialwavephi}
\ee  
By requiring the scalar function~\eqref{eq:partialwavephi} to obey the boundary conditions~\eqref{eq:inmodes} we may write, in the limit $r\rightarrow \infty$, 
\begin{equation}
\psi_{\omega l}(\rt)\sim e^{-i \omega \rt}+{\cal R}_{\omega l }e^{i \omega \rt}
\label{eq:psi_infinity}
\end{equation}

The current $j^{\mu}$ associated to the scalar waves can be defined as
\be
j^{\mu}\equiv \dfrac{i}{2}\l[\phi^{*}\partial^{\mu}\phi-\phi\,\partial^{\mu}\phi^{*}\r].
\label{eq:current}
\ee
Moreover, we can define the flux $\mathcal{F}$ through a surface $\Sigma$ with radius r, as
\be
\mathcal{F} = -\int_{\Sigma}\sqrt{-g}\,j^{r}\text{d}\Sigma,
\label{eq:flux}
\ee 
where $\sqrt{-g}\,\text{d}\Sigma=r^2\sin\theta\text{d}\theta\text{d}\phi$, in spherical coordinates. 

Hence, considering an incident plane wave along the $z$-axis, $e^{-i\omega(t - z)}$, the modulus of the incident current is
\be
|j_{in}|=\omega.
\label{eq:currentin}
\ee
Moreover, substituting Eq.~\eqref{eq:partialwavephi} into Eq.~\eqref{eq:flux}, and using the orthogonality of the Legendre polynomials, namely
\be
\int\sin\theta\text{d}\theta P_{m}(\cos\theta)P_{n}(\cos\theta) = \dfrac{2}{2m+1}\delta_{m n}, 
\label{eq:legpolynorm}
\ee 
we obtain, for the asymptotic radial function~\eqref{eq:psi_infinity}, that
\be
\mathcal{F}=\dfrac{\pi}{\omega}\sum^{\infty}_{l=0}(2l+1)\l(1-|{\cal R}_{\omega l }|^2\r).
\label{eq:fluxphi}
\ee

One can define the (total) absorption cross section $\sigma$ as the ratio between the flux going through the BH event horizon and the modulus of the incident current, namely
\be
\sigma \equiv \dfrac{\mathcal{F}}{|j_{in}|}.
\label{eq:absdef}
\ee
By substituting Eqs.~\eqref{eq:currentin} and~\eqref{eq:fluxphi} in Eq.~\eqref{eq:absdef}, and using the relation~\eqref{eq:RTrelation}, we obtain
\be
\sigma = \sum^{\infty}_{l=0}\sigma_{l},
\label{eq:totabs}
\ee   
where the $\sigma_{l}$ are the partial absorption cross sections, given by
\be
\sigma_{l} \equiv \dfrac{\pi}{\omega^2}(2l+1)|{\cal T}_{\omega l }|^2.
\label{eq:partialabs}
\ee

The absorption cross section has been studied in literature for a vast number of stationary BHs~\cite{fabbri,unruh,sanchez,BODC:2014,MC:2014,CDHO:2015,BC:2016,LDC:2017,LDC:2018,LBC:2019}. The low-frequency behavior of the absorption of scalar fields by stationary BHs is well-known. In the zero-frequency limit the absorption cross section goes to the surface area of the event horizon~\cite{unruh,das,higuchi}. The high-frequency behavior of the scalar absorption has also been extensively studied, culminating in a nice analytical result in this regime, known as sinc approximation~\cite{sanchez}. D{\'e}canini et al. obtained, for an arbitrary static spherically symmetric BH, in the high-frequency regime, the following analytical approximation,~\cite{folacci}
\begin{equation}
\sigma^{hf}_{abs} = \sigma_{geo}\left[1-8\pi\beta e^{-\pi\beta}\text{sinc}\Big(\dfrac{2\pi\omega}{\Omega_{0}}\Big)\right],
\label{eq:sinc}
\end{equation}
where $\text{sinc}\,z=\sin z/z$ is the sine cardinal. The $\sigma_{geo}$ is the geometrical absorption cross section, that can be obtained studying the null geodesics in SKZBH spacetimes. The $\beta$ factor is related to the Lyapunov exponent at the unstable circular orbits radius, $\Lambda_{c}$, and the orbital frequency, $\Omega_{0}$, by~\cite{raffaelli}
\begin{equation}
\beta=\dfrac{\Lambda_{c}}{\Omega_{0}}.
\label{eq:lyapunov}
\end{equation}

\subsection{Numerical method}\label{sec:absorption1}

In order to obtain the coefficients ${\cal T}_{\omega l }$, we solve Eq.~\eqref{eq:radialOHS1} numerically, from near the event horizon to a sufficiently large radius $r_{\inf}$, imposing the boundary conditions~\eqref{eq:inmodes}. We match the numerical solution $\psi^{\text{num}}_{\omega l}$ and its derivative $\text{d}\psi^{\text{num}}_{\omega l}/\text{d}r$ with the expansion at the numerical infinity $r_{\inf}$, obtaining a linear system, namely

\be
\l\{
\begin{array}{ll}
	\psi^{\text{num}}_{\omega l}\Big\vert_{r=r_{\text{inf}}} &=\dfrac{1}{{\cal T}_{\omega l }}e^{-i\omega r_{\text{inf}}} + \dfrac{{\cal R}_{\omega l }}{{\cal T}_{\omega l }}e^{i\omega r_{\text{inf}}}\\
	\dfrac{\text{d}\psi^{\text{num}}_{\omega l}}{\text{d}r}\Big\vert_{r=r_{\text{inf}}} &=-i\omega\l(\dfrac{1}{{\cal T}_{\omega l }}e^{-i\omega r_{\text{inf}}} - \dfrac{{\cal R}_{\omega l }}{{\cal T}_{\omega l }}e^{i\omega r_{\text{inf}}}\r). 
\end{array}\r.\label{eq:numericalmatch}
\ee
By solving the linear system~\eqref{eq:numericalmatch}, we obtain the numerical value of ${\cal T}_{\omega l }$, which we use in Eq.~\eqref{eq:partialabs} to compute the partial absorption cross section, and then in Eq.~\eqref{eq:totabs} to obtain the total absorption cross section.
\subsection{Results}\label{sec:absorption2}
\begin{figure}
\centering
\includegraphics[width=3.0in]{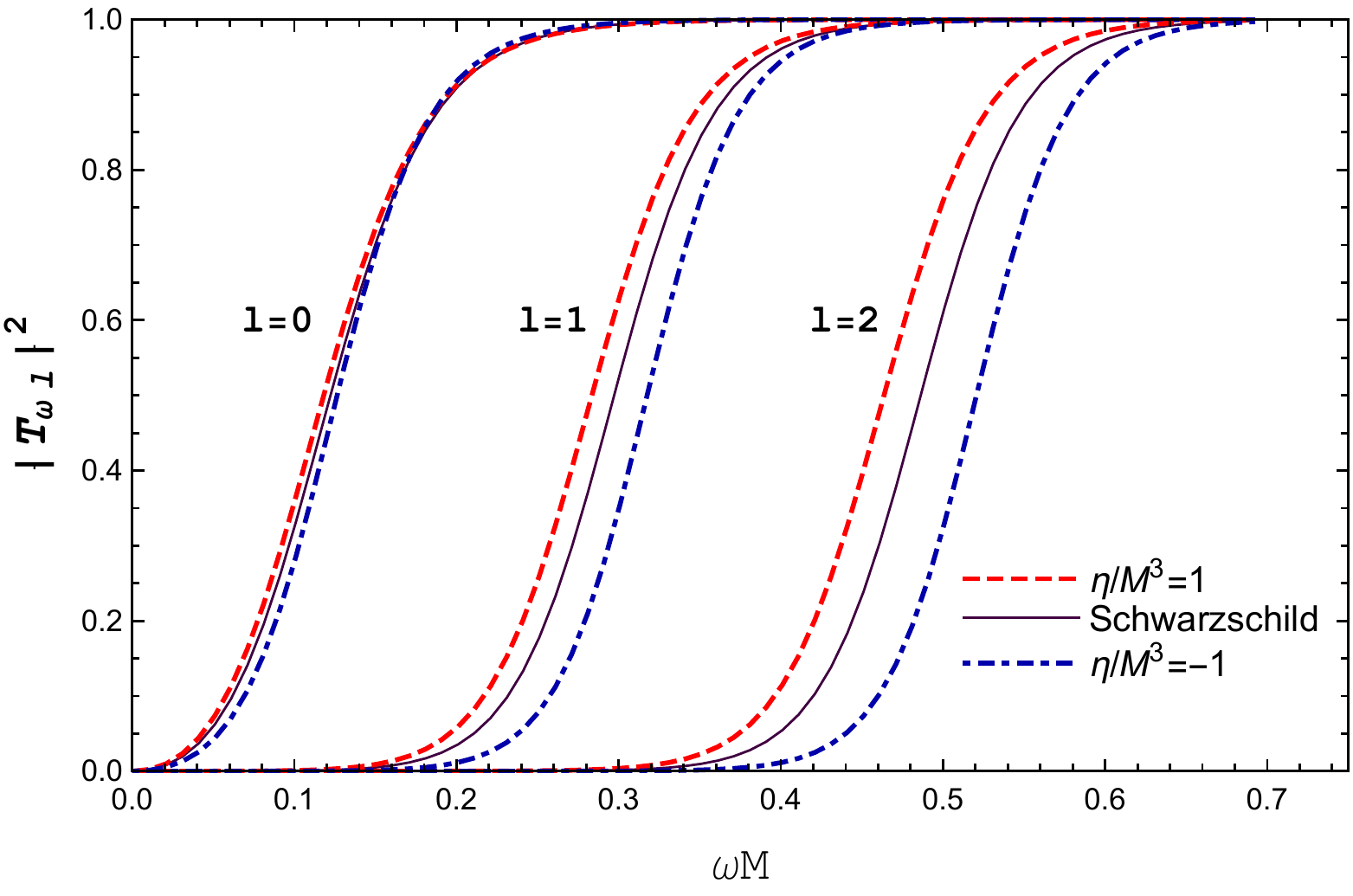}
\caption{The transmission coefficient for some values of the deformation parameter $\eta$, for different $l$-modes.}
\label{fig:transm}
\end{figure}

In Fig.~\ref{fig:transm} we plot the numerical transmission coefficient for some $l$-modes. We notice that, as we increase the deformation parameter $\eta$, the transmission coefficients increase, for all values of $l$. We point out that in the low-frequency regime the only significant contribution to the absorption cross section comes from the $l=0$ mode. 

\begin{figure}
\centering
\includegraphics[width=3.0in]{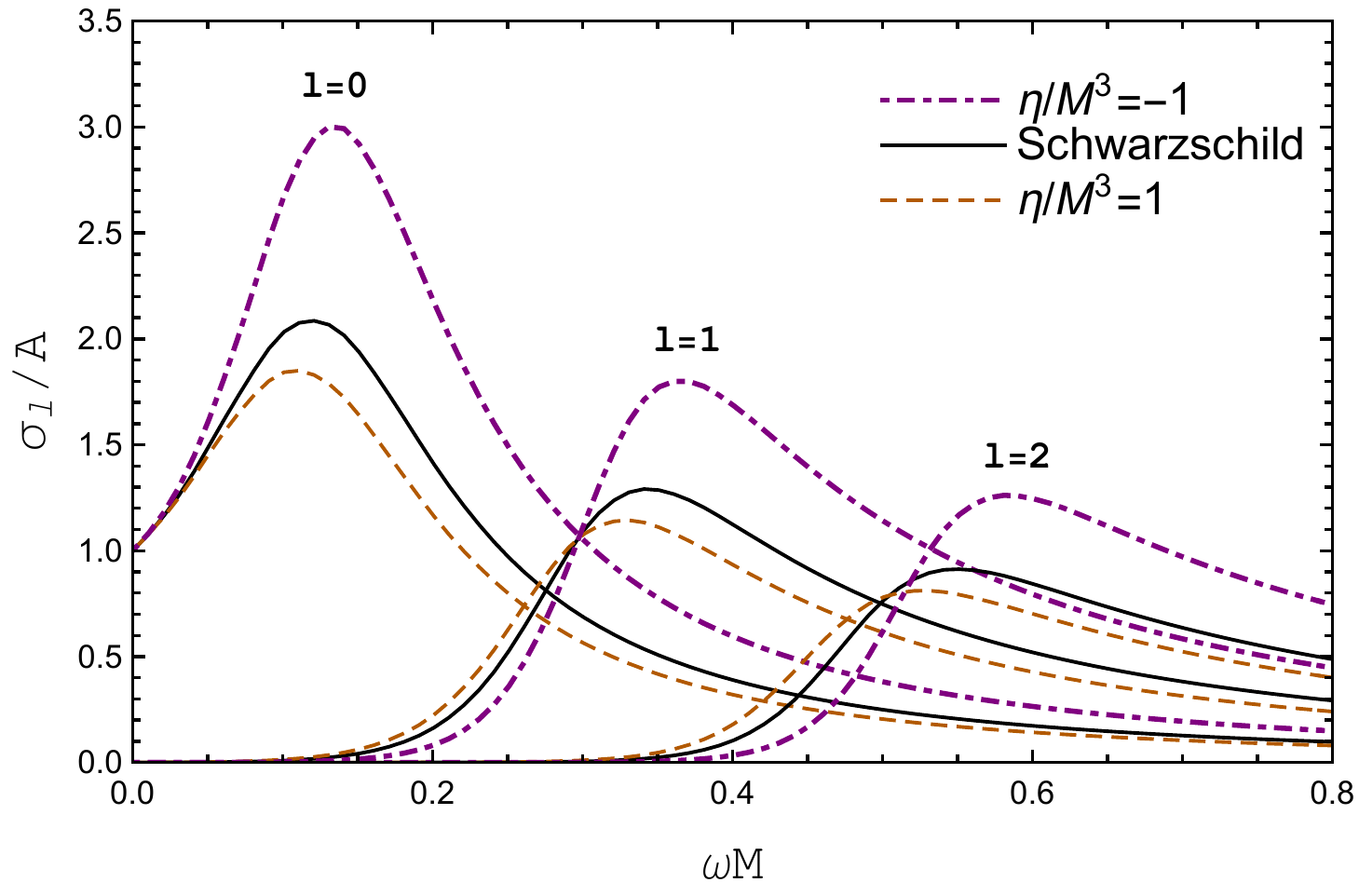}
\caption{The partial absorption cross section, divided by the BH area, for some values of the deformation parameter $\eta$.}
\label{fig:partial}
\end{figure}
We plot in Fig.~\ref{fig:partial}, for some values of $\eta$, the partial absorption cross section. 
We notice that, in the zero-frequency limit, the absorption cross section is essentially given by $\sigma_{0}$, and goes to the area of the BH, in agreement with the general result obtained by Higuchi~\cite{higuchi}.

\begin{figure}
\centering
\includegraphics[width=3.0in]{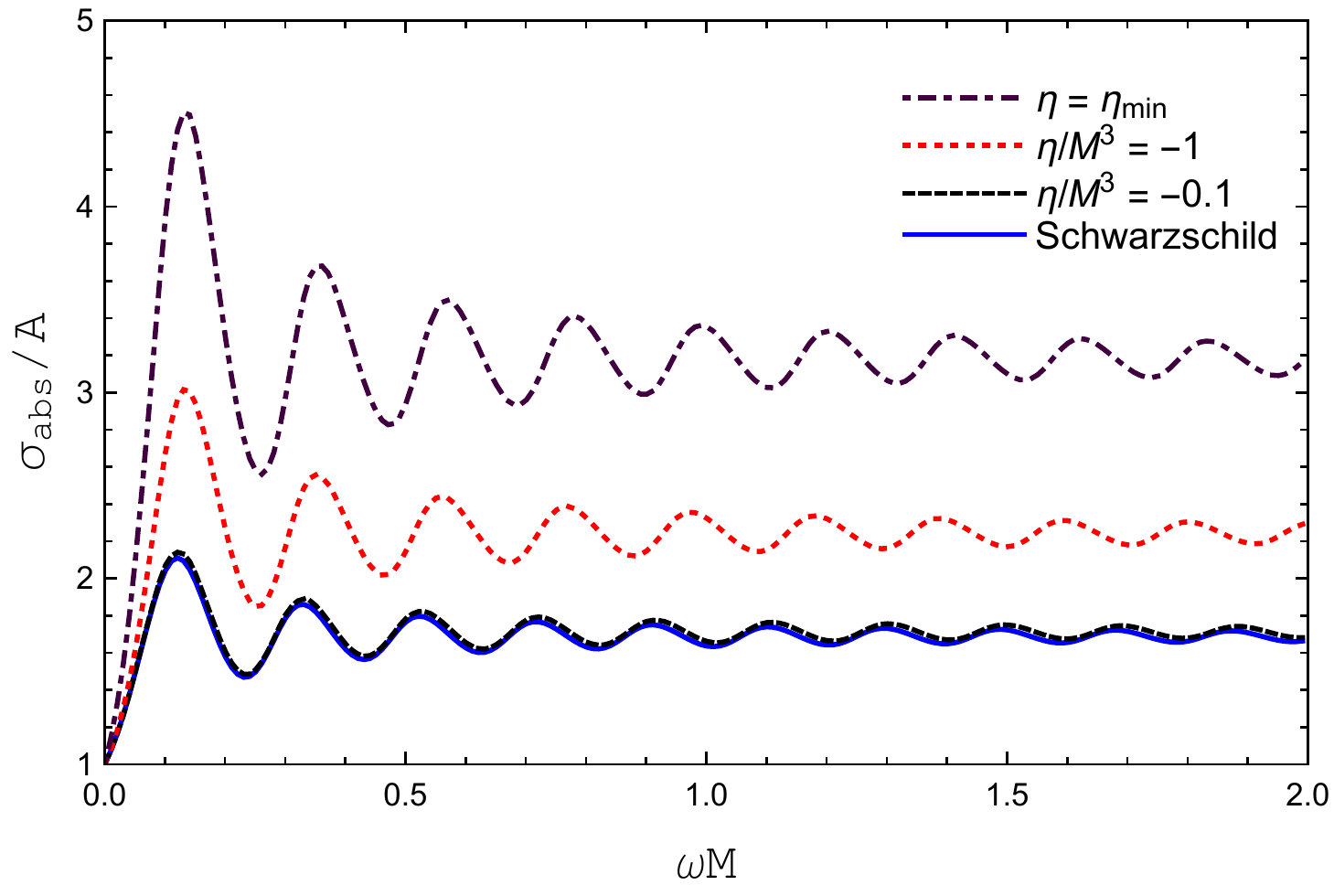}
\includegraphics[width=3.0in]{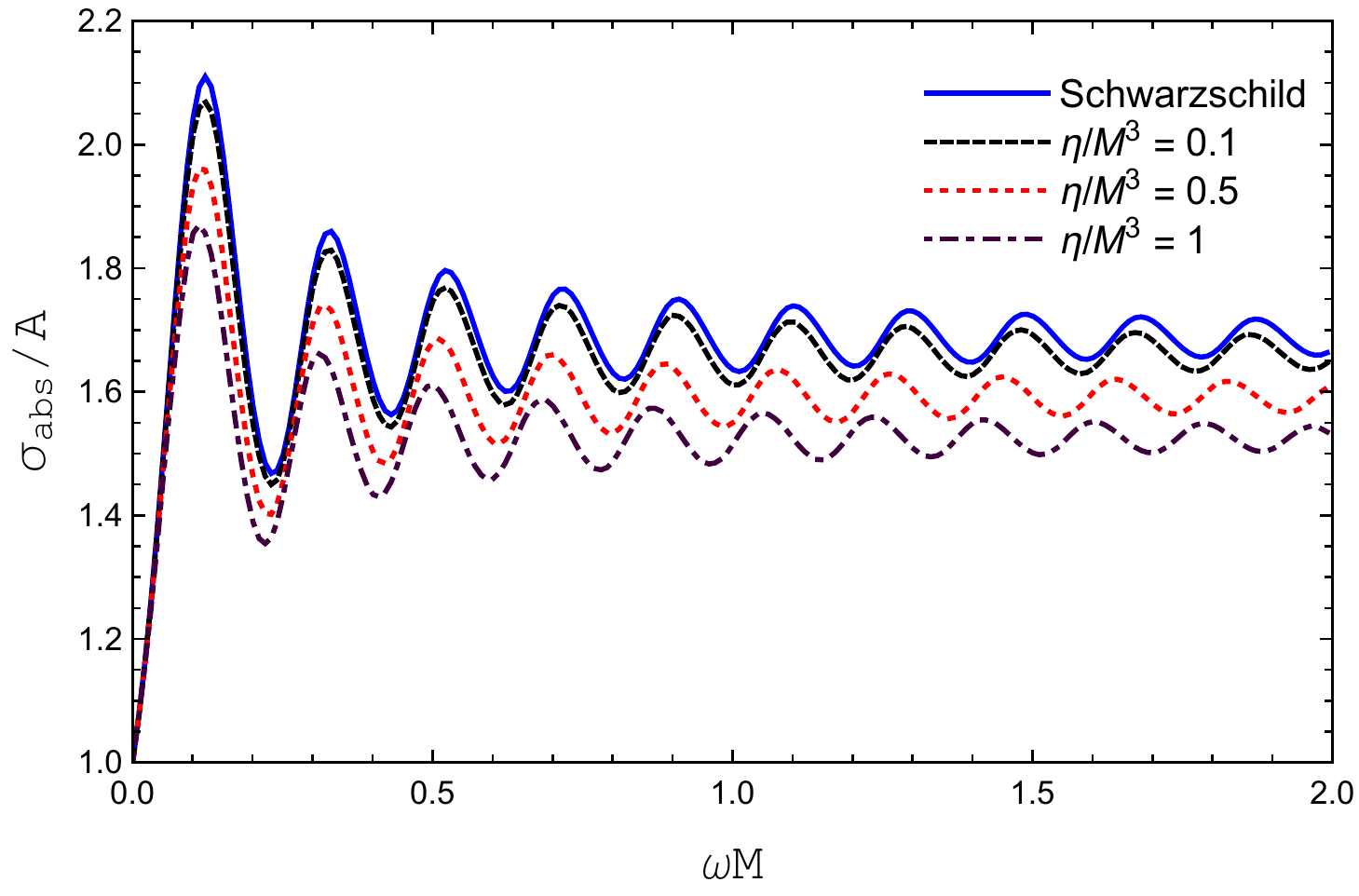}
\caption{The total absorption cross section normalized by the area of the correspondent SKZBH, for non-positive (top panel) and non-negative (bottom panel) values of the deformation parameter $\eta$. We notice that in the low-frequency regime, the total absorption cross section goes to the area of the BH.}
\label{fig:normalizedA1}
\end{figure}

In Fig.~\ref{fig:normalizedA1}, we plot the total absorption cross section for non-positive (top panel) and non-negative (bottom panel) values of the KZ deformation parameter $\eta$, and notice that for bigger values of $\eta$ the total absorption, divided by the area of the BH, is smaller. 

\begin{figure}
\centering
\includegraphics[width=3.0in]{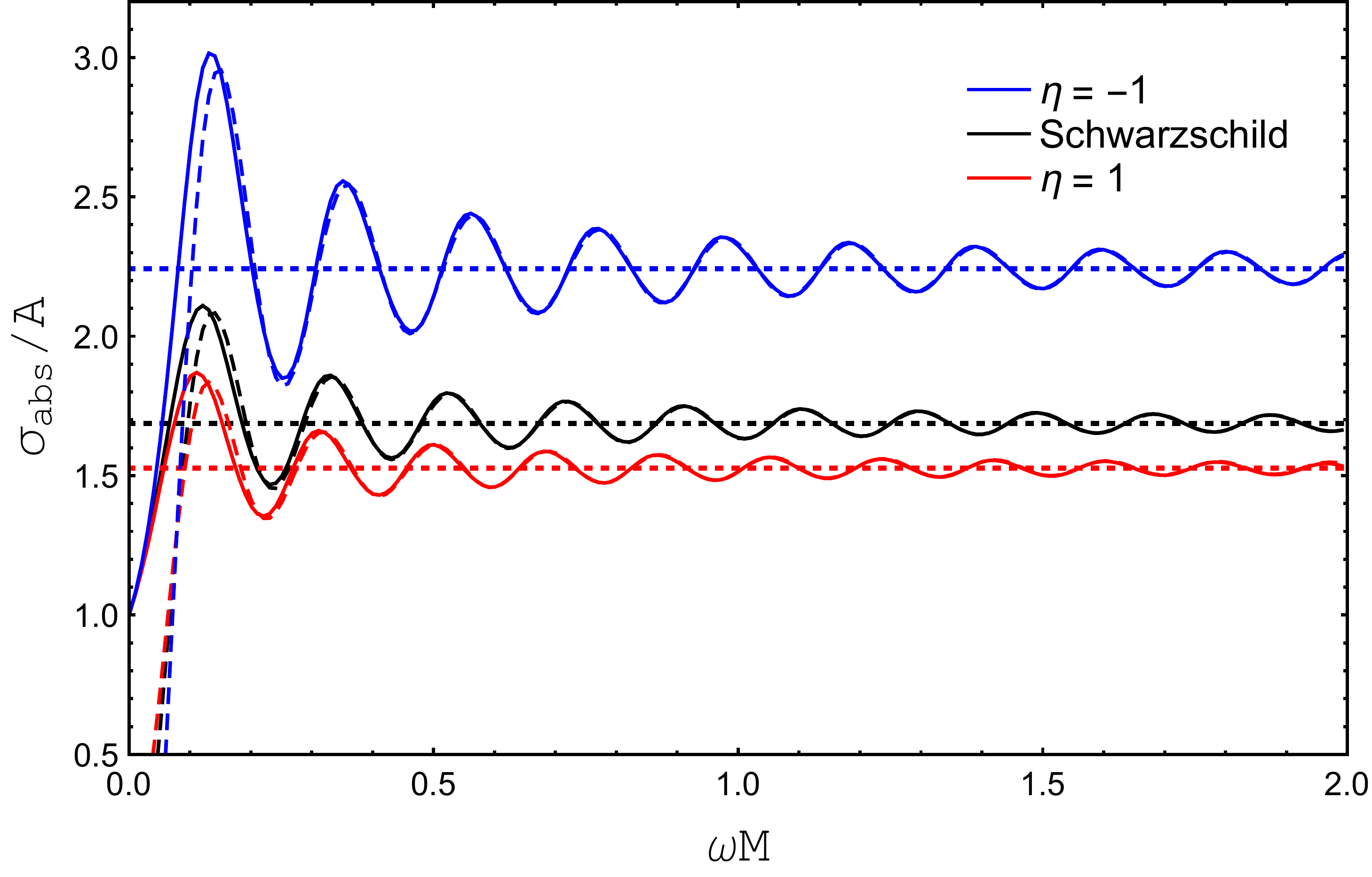}
\caption{The comparison of the numerical absorption cross section (solid curves) with the sinc approximation (dashed curves), for some values of the deformation parameter $\eta$. The horizontal dotted lines represent the geometrical cross section of each corresponding value of $\eta$.}
\label{fig:absandsinc}
\end{figure}

We plot in Fig.~\ref{fig:absandsinc} the comparison between the total absorption cross section~\eqref{eq:totabs} and the sinc approximation~\eqref{eq:sinc}, for $\eta/M^3=1$, $0$ and $-1$. 
We notice that the numerical solutions agree very well with the sinc approximations in the mid-to-high-frequency limit, oscillating around the corresponding geometrical absorption cross sections which are also plotted.

\section{Final remarks}\label{sec:final}

There are accumulated evidences that BHs in Nature are rotating, being described by, at least, one parameter beyond their mass: the BH angular momentum. Different parametrizations have been proposed to describe rotating BHs in alternative theories of gravity. We studied the SKZBH as a first approach to understand how the additional KZ deformation parameter influences the structure of the spacetime and its absorption spectrum.
We started reviewing the SKZBH, constructed introducing a deformation in the mass term of the Schwarzschild solution. This deformation is such that the asymptotic behavior is the same as the one of the Schwarzschild BH, but with a different location of the event horizon, so that the area of the SKZBHs varies with the deformation parameter. We analyzed the absorption of scalar waves by SKZBHs. We computed numerically the total absorption cross section of a spin-0 field for SKZBHs and compared it with the behavior found for the well-known Schwarzschild case. We have shown that the KZ additional parameter changes the scalar absorption, decreasing the absorption cross section divided by the correspondent event horizon area, as the value of the deformation parameter is increased. We obtained that the total absorption cross section goes to the area of the BH in the zero-frequency limit, and in the high-frequency regime the absorption cross section oscillates around the geometrical absorption cross section.

\section*{Acknowledgments}

The authors would like to acknowledge 
Conselho Nacional de Desenvolvimento Cient\'ifico e Tecnol\'ogico (CNPq)
and Coordena\c{c}\~ao de Aperfei\c{c}oamento de Pessoal de N\'ivel Superior (CAPES) -- Finance Code 001, from Brazil, for partial financial support. This research has also received funding from the European Union's Horizon 2020 research and innovation programme under the H2020-MSCA-RISE-2017 Grant No. FunFiCO-777740.
\pagebreak


\begin{thebibliography}{00}
\bibitem{israel} W. Israel, ``Event Horizons in Static Vacuum Space-Times'', Phys. Rev. \textbf{164}, 1776 (1967).
\bibitem{carter} B. Carter, ``Axisymmetric Black Hole Has Only Two Degrees of Freedom'', Phys. Rev. Lett. \textbf{26}, 331 (1971).
\bibitem{robinson} D. C. Robinson, ``Uniqueness of the Kerr Black Hole'', Phys. Rev. Lett. \textbf{34}, 905 (1975).
\bibitem{gossan} S. Gossan, J. Veitch, and B. S. Sathyaprakash, ``Bayesian model selection for testing the no-hair theorem with black hole ringdowns'', Phys. Rev. D
\textbf{85}, 124056 (2012).
\bibitem{meidan} J. Meidam, M. Agathos, C. Van Den Broeck, J. Veitch, and
B. S. Sathyaprakash, ``Testing the no-hair theorem with black hole ringdowns using TIGER'', Phys. Rev. D \textbf{90}, 064009 (2014).
\bibitem{isi} M. Isi, M. Giesler, W. M. Farr, M. A. Scheel, and S. A. Teukolsky, ``Testing the No-Hair Theorem with GW150914'', Phys. Rev. Lett. \textbf{123}, 111102 (2019).
\bibitem{eht} K. Akiyama et al. (Event Horizon Telescope Collaboration), ``First M87 Event Horizon Telescope Results. VI. The Shadow and Mass of the Central Black Hole'',
Astrophys. J. \textbf{875}, L6 (2019); ``First M87 Event Horizon Telescope Results. II. Array and Instrumentation'', \textbf{875}, L2 (2019); ``First M87 Event Horizon Telescope Results. III. Data Processing and Calibration'', \textbf{875}, L3
(2019); ``First M87 Event Horizon Telescope Results. IV. Imaging the Central Supermassive Black Hole'', \textbf{875}, L4 (2019); ``First M87 Event Horizon Telescope Results. V. Physical Origin of the Asymmetric Ring'', \textbf{875}, L5 (2019).
\bibitem{ligo} B. P. Abbott et al. (LIGO Scientific and Virgo Collaborations), ``GW151226: Observation of Gravitational Waves from a 22-Solar-Mass Binary Black Hole Coalescence'',
Phys. Rev. Lett. \textbf{116}, 241103 (2016); ``Observation of Gravitational Waves from a Binary Black Hole Merger'', \textbf{116}, 061102 (2016).
\bibitem{johannsen}	T. Johannsen and D. Psaltis, ``Metric for rapidly spinning black holes suitable for strong-field tests of the no-hair theorem'', Phys. Rev. D \textbf{83}, 124015 (2011).
\bibitem{collins} N. A. Collins and S. A. Hughes, ``Towards a formalism for mapping the spacetimes of massive compact objects: Bumpy black holes and their orbits'', Phys. Rev. D \textbf{69}, 124022
(2004).
\bibitem{vigeland} S. J. Vigeland and S. A. Hughes, ``Spacetime and orbits of bumpy black holes'', Phys. Rev. D \textbf{81}, 024030
(2010).
\bibitem{yunes} S. Vigeland, N. Yunes, and L. Stein, ``Bumpy black holes in alternative theories of gravity'', Phys. Rev. D \textbf{83},
104027 (2011).
\bibitem{KZ} R. Konoplya and A. Zhidenko, ``Detection of gravitational waves from black holes: Is there a window for alternative theories?'', Phys. Lett. B \textbf{756}, 350 (2016).
\bibitem{fabbri} R. Fabbri, ``Scattering and absorption of electromagnetic waves by a Schwarzschild black hole'', Phys. Rev. D \textbf{12}, 933 (1975).
\bibitem{unruh} W. G. Unruh, ``Absorption cross section of small black holes'', Phys. Rev. D \textbf{14}, 3251 (1976).
\bibitem{sanchez} N. G. Sanchez, ``Absorption and emission spectra of a Schwarzschild black hole'', Phys. Rev. D \textbf{18}, 1030 (1978).
\bibitem{BODC:2014} C. L. Benone, E. S. Oliveira, S. R. Dolan, and L. C. B. Crispino, ``Absorption of a massive scalar field by a charged black hole'', Phys. Rev. D \textbf{89}, 104053 (2014).
\bibitem{MC:2014} C. F. B. Macedo and L. C. B. Crispino, ``Absorption of planar massless scalar waves by Bardeen regular black holes'', Phys. Rev. D \textbf{90}, 064001 (2014).
\bibitem{CDHO:2015} L. C. B. Crispino, S. R. Dolan, A. Higuchi, and E. S. Oliveira, ``Scattering from charged black holes and supergravity'', Phys. Rev. D \textbf{92}, 084056 (2015).
\bibitem{BC:2016} C. L. Benone and L. C. B. Crispino, ``Superradiance in static black hole spacetimes'', Phys. Rev. D \textbf{93}, 024028 (2016).
\bibitem{LDC:2017} L. C. S. Leite, S. R. Dolan, and L. C. B. Crispino, ``Absorption of electromagnetic and gravitational waves by Kerr black holes'', Phys. Lett. B \textbf{774}, 130 (2017).
\bibitem{LDC:2018} L. C. S. Leite, S. R. Dolan, and L. C. B. Crispino, ``Absorption of electromagnetic plane waves by rotating black holes'', Phys. Rev. D \textbf{98}, 024046 (2018).
\bibitem{LBC:2019} L. C. S. Leite, C. L. Benone, and L. C. B. Crispino, ``On-axis scattering of scalar fields by charged rotating black holes'', Phys. Lett. B \textbf{795}, 496 (2019).
\bibitem{PPN} C. M. Will, ``The Confrontation between General Relativity and Experiment'', Living Rev. Rel. \textbf{9}, 3 (2006).
\bibitem{bertotti} B. Bertotti, L. Iess, and P. Tortora, ``A test of general relativity using radio links with the Cassini spacecraft'', Nature (London) \textbf{425}, 374 (2003).
\bibitem{boggs} J. G. Williams, S. G. Turyshev, and D. H. Broggs, ``Progress in Lunar Laser Ranging Tests of Relativistic Gravity'', Phys. Rev. Lett. \textbf{93}, 261101 (2004).		
\bibitem{long} F. Long, S. Chen, S. Wang, J. Jing, ``Energy extraction from a Konoplya–Zhidenko rotating non-Kerr black hole'', Nucl. Phys. B \textbf{926}, 83 (2018). 
\bibitem{penrose} R. Penrose, ``Gravitational Collapse: The Role of General
Relativity'', Riv. del Nouvo Cimento, \textbf{1}, 252 (1969).
\bibitem{zwillinger} D. Zwillinger, {\it Handbook of Differential Equations} (MA: Academic Press, Boston, 1997).
\bibitem{das} S. R. Das, G. Gibbons, and S. D. Mathur, ``Universality of Low Energy Absorption Cross Sections for Black Holes'', Phys. Rev. Lett. \textbf{76}, 417 (1997).
\bibitem{higuchi} A. Higuchi, ``Low-frequency scalar absorption cross sections for stationary black holes'', Classical Quantum Gravity \textbf{18},
L139 (2001); ``Addendum to 'Low-frequency scalar absorption cross sections for stationary black holes'', Classical Quantum Gravity \textbf{19},
599(A) (2002).
\bibitem{folacci} Y. D{\'e}canini, G. Esposito-Far{\`e}se, and A. Folacci, ``Universality of high-energy absorption cross sections for black holes'', Phys. Rev. D \textbf{83}, 044032 (2011).
\bibitem{raffaelli} Y. Décanini, A. Folacci, and B. Raffaelli, ``Unstable circular null geodesics of static spherically symmetric black holes, Regge poles, and quasinormal frequencies'', Phys. Rev. D \textbf{81}, 104039 (2010).

\end{thebibliography}
\end{document}